\documentclass[twocolumn,showpacs,nofootinbib,preprintnumbers,amsmath,amssymb,a4paper,superscriptaddress]{revtex4-1}
\usepackage{array}
\usepackage{color}
\usepackage{dcolumn}
\usepackage{bm}
\usepackage{epsf}
\usepackage[dvips]{graphicx}
\include{newcom}
\newcommand{\beq}{\begin{equation}}
\newcommand{\eeq}{\end{equation}\noindent}
\newcommand{\bean}{\begin{eqnarray*}}
\newcommand{\eean}{\end{eqnarray*}\noindent}
\newcommand{\bea}{\begin{eqnarray}}
\newcommand{\eea}{\end{eqnarray}\noindent}

\begin{document}

\bibliographystyle{apsrev}

\title{Search for new physics with neutrinos at Radioactive Ion Beam facilities}

\author{Catalina Espinoza}

\email{espinoza@ipno.in2p3.fr}

\affiliation{Centro de F{\'{\i}}sica Te\'orica de Part{\'{\i}}culas,
Instituto Superior T\'ecnico, Universidade T\'ecnica de Lisboa,
Av. Rovisco Pais 1, 1049-001 Lisboa, Portugal.}

\affiliation{Departamento de F{\'{i}}sica Te\'orica and IFIC, Universidad de
Valencia-CSIC, E-46100, Valencia, Spain.}

\author{Rimantas Lazauskas}

\email{Rimantas.Lazauskas@IReS.in2p3.fr}

\affiliation{IPHC, IN2P3-CNRS/Universit\'e Louis-Pasteur BP 28, F-67037
Strasbourg cedex 2,France.}

\author{Cristina Volpe}

\email{volpe@ipno.in2p3.fr}

\affiliation{AstroParticule et Cosmologie (APC), Universit\'e Paris Diderot - Paris 7, 10, rue Alice Domon et L\'eonie Duquet, 75205 Paris cedex 13, France.}

\date{\today}


\begin{abstract}
We propose applications of Radioactive Ion Beam facilities to investigate physics beyond the Standard Model.
In particular, we focus on the possible measurement of coherent neutrino-nucleus
scattering and on a search for sterile neutrinos, by means of a low energy
beta-beam with a Lorentz boost factor $\gamma \approx 1$.
In the considered setup the collected radioactive ions  are sent inside a 4$\pi$
detector. For the first application we provide the number of events associated
with neutrino-nucleus coherent scattering, when the detector is filled in with a
noble liquid.
For the sterile search we consider that the spherical detector is filled in with
a liquid scintillator, and that the neutrino detection channel is inverse-beta
decay.  We provide the exclusion curves for the sterile neutrino mixing
parameters, based upon the 3+1 formalism, depending upon the  achievable ion
intensity. Our results are obtained both from 
total rates, and including spectral information 
with binning in energy and in distance. The proposed experiment represents a possible
alternative to clarify the current anomalies observed in neutrino experiments.
\end{abstract}

\maketitle


\section{Introduction} \label{sec:intro}

\noindent
A wealth of experimental results on neutrino oscillations have been gathered
since the neutrino oscillation discovery. Currently most of the data from
accelerators, reactors and the sun are interpreted within the theoretical
framework of three active neutrino flavors involving the
Maki-Nakagawa-Sakata-Pontecorvo matrix relating the flavor to the mass basis.
In this case the number of unknown parameters is limited to three angles and three (one
Dirac and two Majorana) phases, most of which has been determined. Recently, the
T2K collaboration has found an indication for a non-zero value for the still
unknown neutrino mixing angle $\theta_{13}$, at 2.5$\sigma$
\cite{arXiv:1106.2822}. A non-zero $\theta_{13}$ is also consistent with the
first Double-Chooz results \cite{Abe:2011fz}. New results on the
third neutrino mixing angle
have recently been obtained by the Daya-Bay  \cite{An:2012eh} and RENO
\cite{Ahn:2012nd} collaborations. The most precise measurement is currently $\sin^2
2\theta_{13} = 0.092 \pm 0.016(stat) \pm 0.005(syst)$, at 5.2 $\sigma$
from Daya-Bay  \cite{An:2012eh}.
Note that a combined analysis had previously favoured a non-zero
$\theta_{13}$ (see e.g. \cite{Balantekin:2008zm,Fogli:2009ce}).
Beyond the intrinsic theoretical interest of
knowing the last mixing angle value, its determination is a key step for setting
up a strategy to search for leptonic CP violation. With an upgrade of the T2K
and NO$\nu$A accelerator experiments, a (small) fraction of the Dirac $\delta$
values can be explored  \cite{Huber:2009cw}. The coverage of most of the Dirac
phase values can be attained only with next generation experiments  including
superbeams or beta-beams  (see e.g. \cite{Volpe:2006in}). The DAEDALUS project
constitutes an interesting alternative \cite{Alonso:2010fs}.
Other open questions concerning fundamental neutrino properties include the
neutrino mass scale, for which the KATRIN experiment should deliver results in
the coming years \cite{Bonn:2010zz}, the neutrino mass hierarchy, the Majorana
versus Dirac nature of neutrinos, and the possible existence of sterile
neutrinos.

 Besides the essential information gathered from terrestrial experiments,
 neutrino properties have an important impact on astrophysical and cosmological observations. 
 Numerous examples exist in the literature showing that information can be extracted on unknown neutrino properties, or discussing their implications on a variety of phenomena
like for example  (stellar and primordial) nucleosynthesis processes.  Recently
it has been shown e.g. that a non-zero CP violating Dirac phase might have an
impact in core-collapse supernovae \cite{Balantekin:2007es,Gava:2008rp}, or on
Big-Bang nucleosynthesis \cite{Gava:2010kz}. Numerous studies have investigated
the effects of sterile neutrinos, e.g. on the r-process (such as
Ref.\cite{McLaughlin:1999pd}) or on the primordial light element abundances,
like in Ref.\cite{Abazajian:2004aj}. Recent cosmological constraints on sterile
neutrinos can be found in \cite{Hamann:2011ge}.

While neutrino oscillations are nowadays an established fact, several anomalies
have recently been observed, that cannot be explained within the standard three
active neutrino framework.
First the MiniBooNE anti-neutrino and neutrino oscillation results are not fully
understood, while an\ increased statistics should help to elucidate the low energy
excess and the oscillation hypothesis \cite{AguilarArevalo:2008rc}. This
experiment, which was supposed to confirm/rule out LSND, has found an
indication for neutrino oscillations at a $\Delta m^2$ of about 1~eV$^2$ both in
the antineutrino channel, using decay-at-rest muons
\cite{Athanassopoulos:1996jb}, and the neutrino channel, based upon
decay-in-flight pions \cite{Athanassopoulos:1997pv}. Note that the KARMEN
experiment employing a similar neutrino source and detector has found no
indication for oscillations and has excluded most of the LSND oscillation
parameter region \cite{Zeitnitz:1998qg}.  The second anomaly is known as the
reactor ``anomaly'' \cite{arXiv:1101.2755}. Indeed a recent reevaluation of the
electron anti-neutrino flux from reactors has shown a shift in the flux
renormalization by 3$\%$ \cite{arXiv:1101.2663}, compared to the previous
predictions. The reanalysis of the reactor experiments, using this new flux, has
shown a significant inconsistency with the three neutrino oscillation
hypothesis.
Finally,  some years ago, the GALLEX and SAGE experiments pointed out an anomaly
in the neutrino flux measured by putting an intense static $^{37}$Ar and
$^{51}$Cr sources inside their detectors. This is referred to as the Gallium anomaly.
Ref.\cite{Giunti:2010zu} has performed a detailed analysis including the
5-10$\%$ uncertainty on the corresponding neutrino-nucleus cross sections, showing that the
Gallium anomaly statistical significance is at the level of 3$\sigma$.

Currently the ensemble of the accelerator, reactor and Gallium anomalies are the
object of debate and have triggered an intense investigation. The possible
interpretations exploit for example one or more sterile neutrinos, such as in
\cite{arXiv:1107.1452}, a combination of sterile neutrinos with non-standard
interactions like in Ref.\cite{Akhmedov:2010vy}, while none of the proposed explanations so
far provides a comprehensive understanding of all the data. Numerous
proposals are being put forward to confirm/rule out possibilities
\cite{arXiv:1107.2335,Vergados:2011na,Anderson:2012pn,Dwyer:2011xs,Yasuda:2011np}. Among these,
Ref.\cite{arXiv:1107.2335} proposes to put intense radioactive sources inside
neutrino detectors, while  Ref.\cite{Agarwalla:2010gd} has pointed out
the possibility to use intense ion sources produced at nuclear facilities. It is clear that independent and aimed experiments
are necessary, to clarify the present situation.

Ref.\cite{Volpe:2003fi} has proposed the idea of
establishing
a low energy beta-beam\footnote{The beta-beam concept was first proposed by Zucchelli to establish a facility for the search of leptonic CP violation \cite{Zucchelli:2002sa}. For a discussion on the status of the feasibility of beta-beam facilities see e.g. \cite{870766}. Note that Ref.\cite{Rubbia:2006pi} has proposed a method to reach high $Q$-value ion intensities e.g. for $^8$B and $^8$Li, which is currently being investigated. } facility, to dispose of neutrino beams in the 100 MeV
energy range, based upon the beta decay of radioactive ions, with $\gamma
\approx 1$ ($\gamma$ being the Lorentz factor) or with an ion boost $\gamma$ of
typically 2 to 7\footnote{Obviously even larger ion boosts, around 10 or more,
would be of interest. The numbers of 2-7 quoted in the available literature on
low energy beta-beams was figured out to keep, in particular, the storage
ring of small size.}.
In the first case, the neutrino fluxes are those of ions that decay-at-rest;
while, in the second case, beams of variable average
energy are obtained through a boost of the ions.
The advantage of having such a facility is to dispose of pure (in flavor) and
well known electron neutrino (or anti-neutrino) fluxes.
The physical applications cover neutrino-nucleus interaction and fundamental
interactions studies, oscillation searches and core-collapse supernova physics,
as pointed out in \cite{Volpe:2003fi} (for a review see \cite{Volpe:2006in}).
These issues have been investigated in detail in a series of works, including
neutrino-nucleus interaction aspects in
\cite{Serreau:2004kx,McLaughlin:2004va,Lazauskas:2007bs,arXiv:1005.2134}, a
measurement of the neutrino magnetic moment in \cite{McLaughlin:2003yg},
a test of the Conserved-Vector-Current hypothesis \cite{Balantekin:2006ga}, a measurement of the Weinberg angle at
low momentum transfer in
\cite{Balantekin:2005md}, to non-standard interactions in \cite{Barranco:2007tz} and of coherent neutrino-nucleus scattering in
\cite{Bueno:2006yq}, the oscillation towards sterile in \cite{arXiv:0907.3145},
an interpolation method to extract the neutrino fluxes from supernovae in
\cite{Jachowicz:2005ym}. In \cite{Volpe:2005iy} the connection between
neutrino-nucleus interaction and double-beta decay is pointed out in relation
with a low energy beta-beam. Most of these applications are based on stored
boosted ions. In Ref.\cite{McLaughlin:2003yg} we have considered the
configuration with $\gamma \approx 1$, with radioactive ions sent to a target
inside a 4$\pi$ detector for the search of the neutrino magnetic moment. Note
that Ref.\cite{Agarwalla:2010gd} has taken the same configuration for a sterile
neutrino search.

In this paper we consider a low energy beta-beam with $\gamma \approx 1$. We
consider that the ions are  injected into  a target inside a 4$\pi$ detector. The
purpose is to use the resulting very low energy neutrino flux to search for new
physics. Here we explore two applications: the search for sterile neutrinos and
a coherent neutrino-nucleus scattering measurement.
We show that, depending on the ion intensity, a coherent neutrino-nucleus
scattering measurement could be performed. Then, we focus on the search for one
sterile neutrino in a 3+1-neutrino flavor framework and present exclusion
plots for the sterile neutrino mixing parameters.
The manuscript is structured as follows. We present our framework in Section II,
while our numerical results are given in Section III. Section IV is a
conclusion.

\section{General framework}

\subsection{Possible setups and corresponding neutrino fluxes}

\noindent

\begin{figure}[!]
\begin{center}
\mbox{\epsfxsize=4.10cm\epsffile{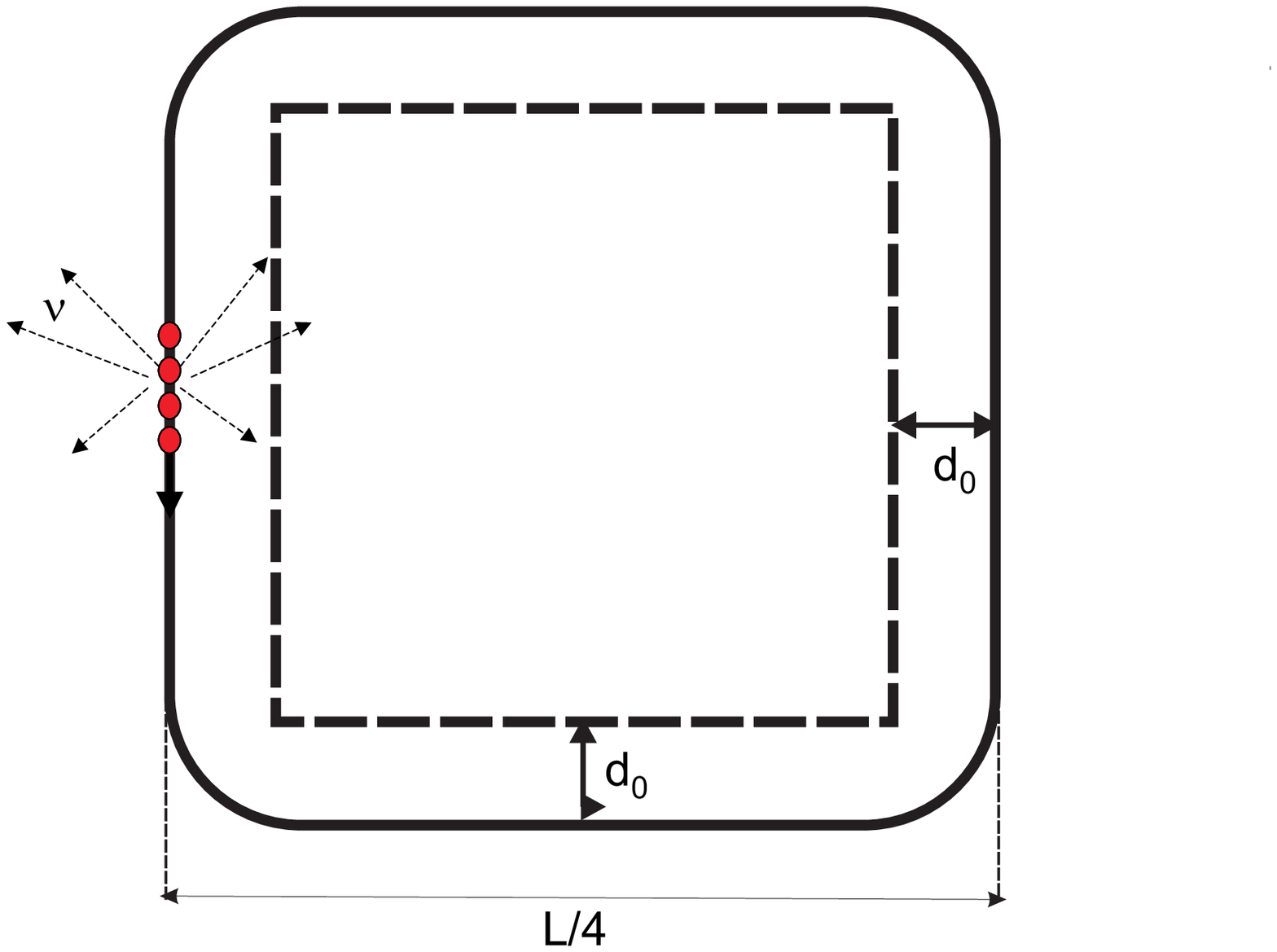}}\hspace{0.1cm}
\mbox{\epsfxsize=3.75cm\epsffile{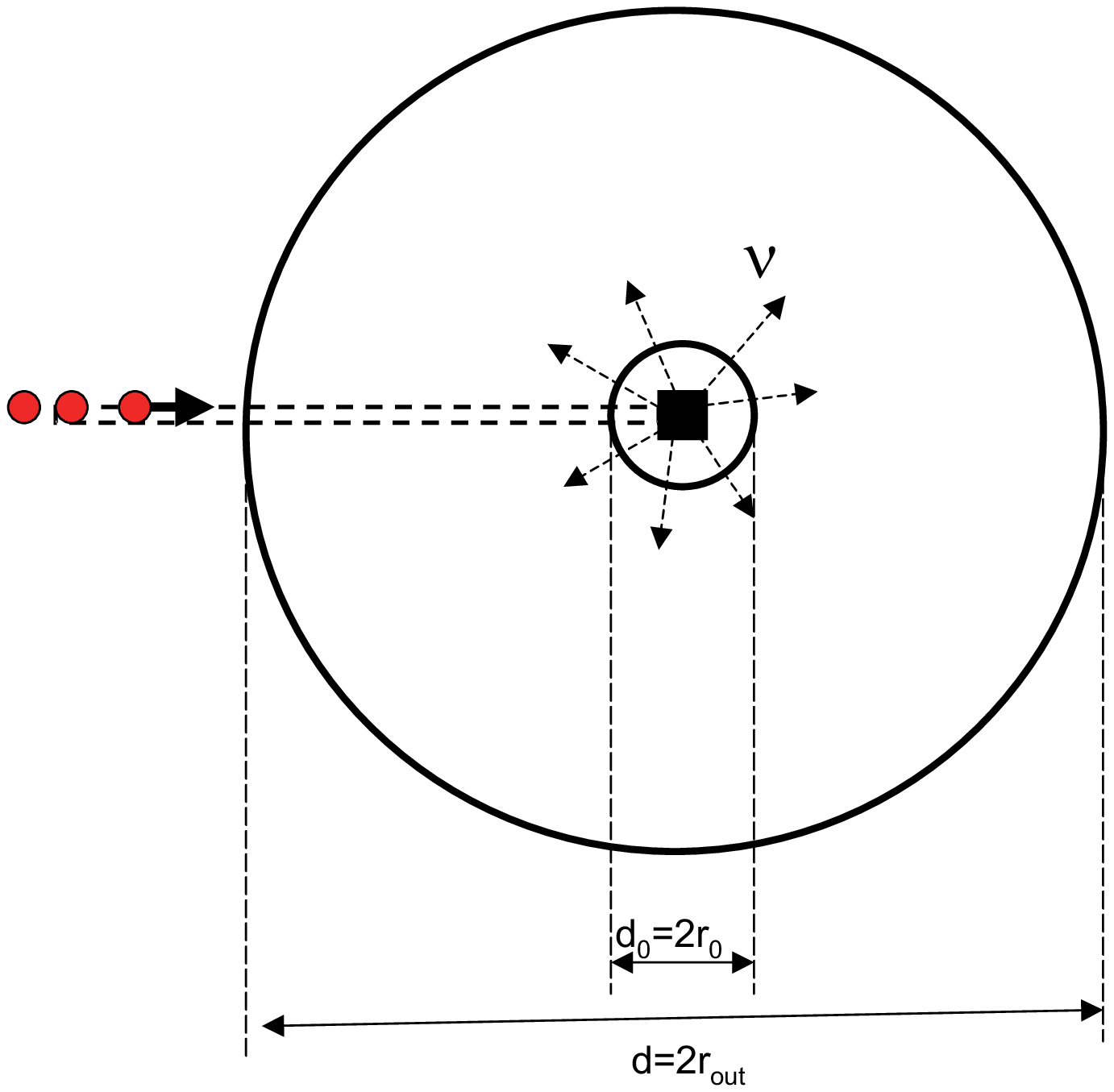}}
\end{center}
\caption{(Color online) Two possible scenarios to produce low energy neutrinos
from  radioactive ions at Lorentz boost of 1. Left figure: The radioactive
ions decay while circulating in a storage ring. The neutrino detector is
installed close to the ring. Right figure:  The radioactive ions are injected
into a cavity inside a spherical detector. }
\label{fig:setup}
\end{figure}

Radioactive ion beam facilities produce intense radioactive ions
decaying through beta-decay, or electron capture. Since specific radioactive ions
can
be selected, a pure electron (anti)neutrino flux can be obtained.
As first proposed in \cite{Volpe:2003fi}, there exists two alternative ways to
produce low energy neutrinos (Figure~\ref{fig:setup}). In the first
scenario the decaying ions
are stored inside a storage ring, while the emitted (anti)neutrinos are
detected in a detector located close to the storage ring.
If the stored ions are boosted, the corresponding neutrino spectra
have variable energy with the average energy given by
$\langle E_{\nu}   \rangle \approx  \gamma Q/2$, with $ Q$ being the $Q$-value of the beta-decaying nucleus. Depending on the application envisaged, the neutrino fluxes can be tuned by appropriately choosing the Lorentz boost and a high/low $Q$-value ion.
In the case the ions are not boosted ($\gamma \approx 1$),
 storing the ions in a small storage ring is a possibility as well. An example is furnished by the storage ring facility
currently proposed at HIE-ISOLDE at CERN \cite{TSR}.
While for this specific storage ring the number of stored ions is limited, one
can imagine the establishment
of a small ring at one of the future intense radioactive ion beam facilities,
such as
European EURISOL  \cite{EURISOL}, or the US Facility for Rare Isotope Beams
(FRIB).

The second scenario to produce low energy neutrinos, consists in injecting the
ions into a target placed inside the detector.
It turns out that, as long as radioactive ions are slow
(i.e. not accelerated to Lorentz boost values above 1), such a
scenario is much more efficient than the storage ring case.
This is due to a geometrical effect since, only
 part of the produced (anti)neutrino flux - and not the total flux - traverses the
detector if the ions are stored in a storage ring. The
average neutrino flux at the detector is further reduced, compared to
the injection inside the detector case, if the detector cannot be located very close to the storage ring due to the background
shieldings and other necessary instrumentation.

In the rest frame, the beta-decay of a nucleus produces the following
(anti-)neutrino flux as a  function of neutrino energy

\begin{equation}
\mathcal{N}(E_{\nu })=f^{-1}E_{\nu }^{2}\,E_{e}\,\sqrt{E_{e}^{2}-m_{e}^{2}}%
\,F(Z,E_{e})\,\Theta (E_{e}-m_{e})\,,  \label{e:1}
\end{equation}%
\noindent where $f$ can be deduced from the measured $%
ft_{1/2}$-value. The quantities appearing in the above expression are the
emitted lepton (electron or positron) energy $E_{e}=Q-E_{\nu }$  and the Fermi
function $F(Z,E_{e})$,
which accounts for the Coulomb modification of the spectrum, with $Z$ being the
ion nuclear charge.

We consider that radioactive ions are produced and injected into a
target with an intensity $I$  of ions per second. This target is installed
inside
a cavity located at the center of a spherical detector. The corresponding
(anti)neutrino flux at the distance $r$ from the target is

\begin{equation}
\phi (E_{\nu },r)=I\frac{\mathcal{N}(E_{\nu })}{4\pi r^{2}}.  \label{e:2}
\end{equation}%
\noindent
The neutrino event rate detected on a spherical surface segment of width $dr$,
located at a distance $r$ from the center of the detector, is given by

\begin{equation}\label{e:3}
\frac{dN_{i}}{dtdE_{\nu }}=\phi (E_{\nu },r)\sigma _{i}(E_{\nu })\frac{%
n_{i}N_{A}\rho }{M_{mol}}4\pi r^{2}dr.
\end{equation}%
Here  $\sigma _{i}(E_{\nu })$ is the neutrino capture cross section on the
target material $i$, $M_{mol}$ is the average molar mass of the detector material,
$n_{i}$ the average
number of nuclei of type $i$ per one mole of the detector material, and
$\rho $ is
its density.

\subsection{A coherent neutrino-nucleus scattering measurement}

The measurement of coherent neutrino-nucleus scattering constitute a precision
test of the Standard Model,
including
 the possibility to probe the weak nuclear charge as well as
  various deviations from the SM predictions, 
due to new physics above the weak scale, or presence of the sterile
neutrinos \cite{Anderson:2012pn}.
Coherent neutrino-nucleus scattering is also important in the astrophysical context, e.g. for core-collapse supernova physics.

Several proposals have been made to perform such a measurement, particularly nearby
Spallation Sources \cite{Scholberg:2009ha,Vergados:2009ei}. Here we consider a
setup as shown in Figure \ref{fig:setup} right.
The cross section for electron neutrino (or anti-neutrino) coherent scattering on
a nucleus is given by
\cite{Freedman:1973yd, Horowitz:2003cz}

\bea\label{e:4}
\frac{d\sigma}{dT} = \frac{G_F^2}{4\pi} Q_W^2 M \left( 1 - \frac{MT}{2E_\nu^2}
\right) F(2MT)^2\,
\eea
here $G_F$ is the Fermi constant, $M$ is the nuclear mass, $T$ the nuclear
recoil energy, $F$ the ground state elastic form factor and

\bea\label{e:4b}
Q_W = N - \left( 1 - 4 \sin^2\theta_W \right) Z
\eea
the weak nuclear charge, with $N$ the number of neutrons, $Z$ the number of
protons and $\theta_W$ the weak mixing angle.
For neutrino energies below $50$ MeV the momentum transfer is small and the form
factor is close to unity $F\sim1$.
For the measurement of neutrino-nucleus coherent scattering, different types of
liquids are being discussed (see e.g.  \cite{Scholberg:2009ha}). Here we take
liquid neon as an example.

\subsection{A 3+1 sterile neutrino oscillation experiment}

\noindent

In the present work we consider a sterile neutrino search within the 3+1
framework with three active neutrinos and one additional sterile neutrino.
Besides the usual parameters of the Maki-Nakagawa-Sakata-Pontecorvo matrix,
in this case the oscillation formula depends upon the neutrino mixing angle
$\theta_{new}$ and $ \Delta m^2_{new}$, considered to be much larger than
$\Delta m_{31}^{2}\simeq 2.4 \cdot 10^{-3}$ $eV^{2}$.
Implementing more complex scenarios with extra neutrinos is straightforward.
The electron neutrino survival probability
for $P_{ee}(E_{\nu },r)$ is given by Ref.~\cite{deGouvea:2008qk}

\bea\label{e:5}
P_{ee}(E_{\nu },r)=1-\cos ^{4}\theta _{new}\sin ^{2}\left( 2\theta
_{13}\right) \sin ^{2}\left( \frac{\Delta m_{31}^{2}r}{4E_{\nu }}\right) \notag
\\
-\sin ^{2}\left( 2\theta _{new}\right) \sin ^{2}\left( \frac{\Delta
m_{new}^{2}r}{4E_{\nu }}\right)
\eea
where a baseline of $L<2$ km and neutrino energies $E_\nu>2$ MeV are assumed.
Eqs.(\ref{e:1}-\ref{e:3}) are used to determine the unoscillated number of
events, while Eq.(\ref{e:3}) has to be multiplied by the neutrino survival
probability $P_{ee}$ Eq.(\ref{e:5}) in order to determine the number of oscillated events.

\subsubsection{Statistical analysis and backgrounds }

\noindent

We present sensitivity plots obtained with the following procedure.
We deal with systematic uncertainties inherent to the experimental
setup by implementing the systematics directly into the statistical analysis
by the use of the pull approach (see for instance
\cite{Huber:2002mx,Fogli:2002au}).
Unless otherwise stated we bin our
data in energy as well as in R intervals of equal spacing, R being the distance from the center of the
detector, the $\chi^2$ function being defined as

\begin{equation}\label{e:6}
\chi^2 = \textrm{min}_{\xi, \bar{\lambda}} \left[  2\left(
         \sum_{ij}{ N_{ij}^t - n_{ij}^f - n_{ij}^f
\ln{\frac{N_{ij}^t}{n_{ij}^f}} }  \right)
          + \xi^2 \right],
\end{equation}
where the sum runs over energy and R-bins. As is customary in this type of
analysis a superscript
$t$ is used to denote the predicted number of events $N_{ij}^t$, while a superscript $f$ is employed to denote the
number of events
obtained from the fitting to the simulated data, $n_{ij}^f$. The systematic error
is indicated by $\pi$. It
enters the analysis through the definition:

\begin{equation}\label{e:7}
N_{ij}^t = n_{ij}^t \times (1 + \pi\xi).
\end{equation}
The respective true and fitted number of events, $n_{ij}^t$ and $n_{ij}^f$, are
functions of their
corresponding oscillating parameters, but nonetheless the marginalization in Eq.
(\ref{e:6}) is
performed over the subset of fitted parameters $\bar{\lambda}$ not held fixed
during the fitting,
as well as over the `pull' $\xi$.



The issue of the background levels is an important one. 
First of all we would like to stress that for the spherical detector setup
(Figure \ref{fig:setup}, right) we will not have sizeable beam\footnote{Note that for the storage ring configuration (Figure \ref{fig:setup}, left)
some background might be induced by the daughter nuclei colliding with the storage ring. Such a background can in principle be suppressed by putting an appropriate shielding around the storage ring.}  or implantation related
backgrounds, since the ions implanted on the target are essentially at rest\footnote{Note that the situation here is very different from the one where the ions are boosted at high $\gamma$. In this case there is again no beam related background (a known advantage of the beta-beam concept) while there is radioactivity induced in the storage ring arising, e.g. from collisions of the stable daughter nuclei on the ring. In this case a shielding is necessary to suppress the related backgrounds and has been considered in the previous literature on low energy beta-beams with $\gamma$ larget than 1.}.
From previous experience
with reactor
experiments the primary sources being environmental and geoneutrino backgrounds,
which one could
deal with in a relative simple fashion by implementing an energy threshold
around 4 MeV in the
neutrino energy. The real problem are atmospheric muons since one gets
backgrounds via their
high energy neutrons or their spallation left overs (Li-9). In order to reduce
this kind of
background signal to a tolerable level one needs a detector design similar to the one of, for
instance, Daya Bay
\cite{Guo:2007ug}, where in addition to a rock overburden of at least 98 m (or
260 mwe), a 20 ton liquid
target must be surrounded by a 20 ton gamma catcher and 40 ton of buffer volume.
Under these
conditions the noise levels can be reduced to $6$ events per ton per year.
Therefore, to achieve such a low background event rate for the measurement under
consideration here, it is necessary to locate the experiment well
underground and surround the detector with appropriate shielding. In our
calculations we assume that this can be done, and take as a reference value for
the background, $6$ events per ton per year.

\section{Numerical results} \label{sec:results}

\noindent

To produce low energy neutrinos both $\beta^+$ and $\beta^-$ decaying ions
can be considered as electron neutrino and anti-neutrino emitters respectively\footnote{Note that electron-capture neutrino beams have been considered in \cite{Bernabeu:2005jh}.}.
The choice of the ions depends on the achievable intensities, the
half-lives and $Q$-values. Obviously the half-lives should lie in an appropriate
range between short and long to make experiments feasible, so that typically
half-lives in the 1 s range seem to be a good choice. On the other hand, high
$Q$-values help increasing the total number of events as well as improving the signal-to-background ratio.

Table \ref{tab:1} shows the candidate
ions that we have been considering here, as typical examples. Note that there exist a
number of other promising radioactive ions, such as e.g. $^8$He or $^{12}$N \cite{Autin:2002ms}.
As far as $^8$Li and $^8$B are concerned, they decay mainly into a broad $^8$Be J$^{\pi}=2^+$
excited state, at 3.03 MeV, therefore having  $Q$-values
centered at 13.1 MeV and 15.1 MeV, respectively. Nevertheless, due to the broadness of the final state,
the neutrino spectrum is extended well above the energy associated to the centered
$Q$-value. To evaluate qualitatively this effect, we present
results for two decay modes \footnote{Note that an accurate neutrino spectrum might be obtained by describing $^8$Be
final state continuum, when taking into account the delayed-$\alpha$ spectrum
\cite{Bhattacharya:2002gc}.} :
i) 100\% branching ratio to the $^8$Be  ground state; ii) 100\% branching ratio to a
narrow excited state at 3.03 MeV.
In a real experiment the actual results will fall in-between these two ``extreme" cases.

For the ion intensity, we assume $10^{13}$ ions per second. Instead of taking this parameter as
a tunable number (as sometimes done in the literature),
here we consider values, that can in principle be achievable at next generation radioactive ion beam facilities. The predictions we
present are obtained by taking 1 year = 10$^7$ s and a 100$\%$ efficiency of the detectors.

\begin{table}
\caption{\label{tab:1} Beta-decay properties of the ions considered in our proposal:
$\tau$ is the decay lifetime, $E_\nu^\textrm{max}$ is the end-point energy.}
\begin{ruledtabular}
\begin{tabular}{lllll}
Ion & Decay & Daughter (State) &  $\tau$(ms)  & $E_\nu^\textrm{max}$(MeV)\\
$^{6}_{2}$He & $\beta^-$ & $^{6}_{3}$Li ($1^+,0$) &  $806.7$  & $3.5078$\\
$^{8}_{3}$Li & $\beta^-$ & $^{8}_{4}$Be ($2^+,0$) &  $838$    & $13.103$\\
$^{8}_{3}$Li & $\beta^-$ & $^{8}_{4}$Be ($0^+,0$) &  $838$    & $16.003$\\
$^{8}_{5}$B  & $\beta^+$ & $^{8}_{4}$Be ($2^+,0$) &  $770$    & $15.079$\\
$^{8}_{5}$B  & $\beta^+$ & $^{8}_{4}$Be ($0^+,0$) &  $770$    & $17.979$\\
\end{tabular}
\end{ruledtabular}
\end{table}

\subsection{Expected $\nu$-nucleus coherent scattering events}

\noindent
For the coherent neutrino-nucleus scattering measurement both electron neutrino
($\beta^+$) and anti-neutrino ($\beta^-$) emitters can be used.
We have considered $^{8}$Li and $^{8}$B and their two decay modes (Table
\ref{tab:1}).
We take a 1 ton spherical
liquid-neon detector, where the ions are injected inside
a central cavity having 20 cm radius (Figure \ref{fig:setup} right). While other target nuclei are obviously
possible, liquid neon is taken as an example.
We would like to emphasise that detailed background simulations have already been done e.g. for the CLEAR experiment proposed at the SNS spallation source facility \cite{Scholberg:2009ha}. The shielding envisaged for such a detector
has been shown to reduce backgrounds nearby spallation sources at a negligible
level. We expect that a similar reduction can be reached by putting the detector underground and/or using an appropriate shielding.
However reaching very low nuclear recoils is challenging for background issues (see, for example Figs. 10 and 11, of Ref.\cite{Scholberg:2009ha}) and for technical features, like light quenching (for a discussion see e.g. Ref.\cite{Mei:2007jn}). Although very optimistic, a sensitivity threshold of $10$ keV will be assumed. A higher energy threshold choice can be an option in the kind of proposals discussed here, only if much higher ion intensities can be attained.

\begin{figure}[!]
\begin{center}
\epsfxsize=7.5cm\epsffile{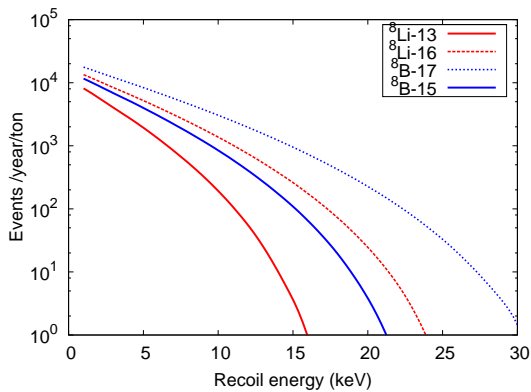}
\end{center}
\caption{ Coherent $\nu$-nucleus scattering: expected
number of events, as a function of the minimal nuclear recoil for the experimental setup of Figure \ref{fig:setup} right. The $^{8}$B and $^{8}$Li ion
intensity is fixed at 10$^{13}$ions/s (here 1 year = 10$^{7}$~s).}
\label{fig:coherentevents}
\end{figure}

Table \ref{tab:II} presents the number of expected events associated to electron
(anti)-neutrino scattering on neon.
One can see that,
despite the fact that coherence enhances the cross-section relative to
other type of processes,
the low energy range covered by the neutrino flux in this work makes the number of events still rather small, compared to the ones attainable with, e.g. the
Michel spectrum of decay-at-rest muons produced at spallation sources
\cite{Scholberg:2009ha}.
Clearly, the feasibility of the measurement we are proposing strongly
depends upon the achievable ion intensities and the lowest measurable nuclear recoil in the detector. Figure \ref{fig:coherentevents} presents our predictions for the number
of events for the
two candidate ions considered, as a function of the minimum measurable nuclear recoil.
Note that it is straightforward to scale our rates (Table \ref{tab:II}) with the ion intensity collected at the center of
the 4$\pi$ detector, or to take into account effects such as light quenching. One can see the strong sensitivity of the results
to the maximal neutrino energy depending on the $Q-$value of the ions.

\begin{table}
\caption{\label{tab:II} Coherent $\nu$-nucleus scattering: expected number of events for the two candidate ions considered and the setup of Figure \ref{fig:setup} right (here 1 year = 10$^{7}$~s).
The maximal neutrino energy is denoted by $E_\nu^\textrm{max}$, the nuclear recoil detection
threshold by T$_\textrm{min}$. The results correspond to
an intensity of $10^{13}$ ions/s.}
\begin{ruledtabular}
\begin{tabular}{llllll}
Ion & Decay & Target & $E_\nu^\textrm{max}$(MeV)  &  T$_\textrm{min}$(keV) &
Events/ton/year\\
$^{8}_{3}$Li & $\beta^-$ & Ne  & $13.103$  &  $10$  &  $192$\\
$^{8}_{3}$Li & $\beta^-$ & Ne  & $16.003$  &  $10$  &  $1373$\\
$^{8}_{5}$B  & $\beta^+$ & Ne  & $15.079$  &  $10$  &  $846$\\
$^{8}_{5}$B  & $\beta^+$ & Ne  & $17.979$  &  $10$  &  $3047$\\
\end{tabular}
\end{ruledtabular}
\end{table}

\subsection{Active-to-sterile neutrino oscillation exclusion plots}

\noindent
For the active-to-sterile neutrino oscillation search, we shall consider two types of $\beta^-$ decaying ions
(Table \ref{tab:1}). First, because of its
very well known aspects relating to its production and management, it is worth
taking a look at
the physics reach of a facility based on $^{6}$He. Its low $Q$-value
yields fewer number of
counts and, as we will show, this hinders the potential of a setup exploiting this ion, instead of one based upon a high $Q$-value ion choice.
Our proposal for the search of sterile 
neutrinos
is mainly based on the properties of
$^{8}$Li, for which we assume two extreme cases, indicated as $^{8}$Li-16MeV and $^{8}$Li-13MeV `ions'.

Our choice of main setup has been dictated by an analysis of the performance of
the two possible
configurations shown in Figure \ref{fig:setup}. In both cases, the considered
detector is filled in with a liquid scintillator\footnote{We take as an example C$_{16}$H$_{18}$ with a density of $\rho=988$ $kg/m^3$.}.
The electron anti-neutrino detection channel is inverse beta-decay
$\bar{\nu}_e + p \rightarrow n+ e^+$.
A good 
signal-to-background ratio can be obtained by the
addition of Gadolinium and the subsequent detection of the 8 MeV prompt
gamma-rays produced by the neutron capture.

For the case of the active-sterile neutrino oscillation hypothesis, under
consideration here,
we have chosen to present the results of our simulations by means of exclusion
plots based upon Eq.(\ref{e:6}).
The exclusion plots for the active-to-sterile oscillation parameter
$\sin^2(2\theta_{\textrm{new}})$ are obtained by
 additionally fixing $\sin^2(2\theta_{13})$\footnote{Note that the
 exclusion curves change little if one fixes the third neutrino mixing angle to zero.}
 to the best fit value of
Ref.\cite{Schwetz:2011zk}, namely $\sin^2( 2\theta_{13} ) = 0.051 $. 
Recently the Daya-Bay collaboration has measured the third neutrino
mixing angle at 5.2 $\sigma$ to be $\sin^2 2\theta_{13} = 0.092 \pm  0.016(stat) \pm 0.005(syst)$ \cite{An:2012eh}.  We have checked that
the plots presented here show no appreciable changes if
the Daya-Bay value is used.
The plots show the oscillation parameter space region
where our setup is expected to be sensitive to the detection of active-sterile neutrino oscillations. 
In all our calculations the considered running time of the experiment is of 5 years.
Unless contrarily stated, we fix the systematic error to
$\pi=1$\%\footnote{Note that it is not our goal to discuss how low
systematic errors can be achieved in the actual experiment. For
example, in Ref.\cite{Giunti:2009en} the authors discuss how this can be done in
a short-baseline experiment at a neutrino factory.} 
in all the analysis
presented hereafter; while we will show how our main results change if
a larger systematic error is considered.
We shall compare the sterile neutrino oscillation parameter regions that can be covered with our experimental setup, to the allowed regions presented in the analysis of Ref. \cite{arXiv:1101.2755}, based on
reactor neutrino experimental data cumulated so far.

\begin{figure}[!]
\begin{center}
\epsfxsize=7.5cm\epsffile{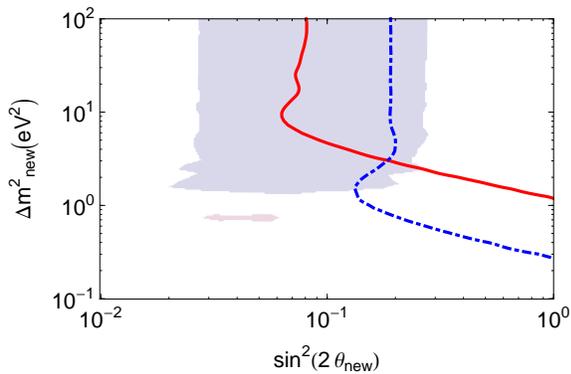}
\end{center}
\caption{Exclusion plots for the sterile neutrino mixing parameters from an analysis of
the data including only total rates. The results are obtained by considering a 3+1 neutrino oscillation formalism.
The contours shown are for a C.L. of 99\% (2 d.o.f.). The two setups are those of Figure \ref{fig:setup}.
The solid (red) line corresponds to the $4\pi$ detector surrounding the source;
while the dashed-dotted (blue) line corresponds to the detector place at the center of the square
storage ring. The ion intensities are of
$10^{11}$ ions/s for the storage ring and of $10^{13}$ ions/s for the
spherical detector (see text). The source is $^{8}$Li ions decaying mainly to the first excited
state of the daughter nucleus (maximal neutrino energy $13$ MeV). In both cases a 5 years running
time is assumed. For comparison the shaded region
represents the 99\% C.L. inclusion domain, given by the combination of reactor neutrino
experiments and other sources (adapted from Fig. 8 of Ref.
\cite{arXiv:1101.2755}).}
\label{fig:comparison}
\end{figure}

Figure \ref{fig:comparison} presents exclusion
plots calculated from a statistical analysis
of the data using total rates.
The facility is based on $^{8}$Li-13 MeV decaying ions.
The aim of the figure is to compare the results obtained for the two experimental setups of Figure \ref{fig:setup}.
Note that for the specific case of the storage ring only,
we assume an intensity of
$10^{11}$ ions/s, having in mind a facility like HIE-ISOLDE (although the stored ion intensity is expected to be smaller \cite{TSR}). Such an intensity should be attainable in a storage ring nearby the EURISOL facility \cite{EURISOL}.
For the setup geometry, following the TSR proposal for HIE-ISOLDE, we take a square storage ring with straight sections of $61.6$ m length and a 1 kton cubic\footnote{The detector base has a size of
$9.4\times9.4$ m and the height is of 11.3 m.
Half of the detector is located below the storage ring, and half above.} detector at the storage ring center.
(Such a geometry leaves 3 meters space between the detector and the storage
ring straight sections \cite{Grieser,Grieser:2012zz}.) Note that, the
number of expected events and, thus the exclusion plots, strongly depends on the setup geometry. For a large
detector, as considered here, placing it in the center of the storage ring represents 
the optimal scenario (if such a detector is located along one storage ring straight section,
the event number is reduced by almost a factor of 4). 
As expected, although the detector is of 20 tons only, the performance obtained by sending the ions inside a 4$\pi$
detector is superior to the storage ring one\footnote{Note that this is also due to the higher ion intensity.} with respect to the coverage of
the shaded region identified by the reactor anomaly. On the other hand,
the storage ring setup has a better sensitivity to small $\Delta m^2$.

\begin{figure}[!]
\begin{center}
\epsfxsize=7.5cm\epsffile{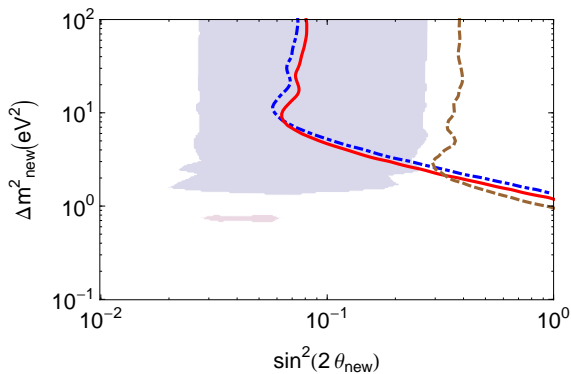}
\end{center}
\caption{Exclusion plots for the sterile neutrino mixing parameters from an analysis of
the data including only total rates. The contours shown are for a C.L. of 99\% (2 d.o.f.).
The results correspond to choosing different ions:
a source of $^{8}$Li ions
decaying mainly to the first excited state of the daughter nucleus (solid red line), $^{8}$Li ions
decaying mainly to the ground state of the daughter nucleus (dashed-dotted blue line), or
a source of $^{6}$He ions (dashed brown line).
For comparison the shaded region
represents the 99\% C.L. inclusion domain, given by the combination of reactor neutrino
experiments and other sources (adapted from Fig. 8 of Ref.
\cite{arXiv:1101.2755}).}
\label{fig:mainU}
\end{figure}

\noindent
In addition to
the aforementioned geometric advantages, the spherical detector setup benefits from the fact that
neutrino source is very close to the active detector material (we remind that the radius cavity is of 20 cm only).
From now on, all the results we present will correspond to the spherical detector setup.

\begin{figure}[!]
\begin{center}
\epsfxsize=7.5cm\epsffile{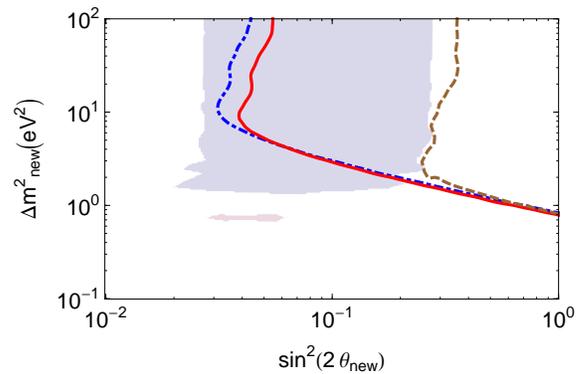}
\end{center}
\caption{Exclusion plots for the sterile neutrino mixing parameters,
with binned analysis of the simulated data. Both binning in neutrino energy and in distance within the detector. The contours shown are for a C.L. of 99\% (2 d.o.f.).
The results correspond to choosing different ions:
a source of $^{8}$Li ions
decaying mainly to the first excited state of the daughter nucleus (solid red line),  $^{8}$Li ions
decaying mainly to the ground state of the daughter nucleus (dashed-dotted blue line), or
a source of $^{6}$He ions (dashed brown line).
For comparison the shaded region
represents the 99\% C.L. inclusion domain, given by the combination of reactor neutrino
experiments and other sources (adapted from Fig. 8 of Ref.
\cite{arXiv:1101.2755}).}
\label{fig:mainB}
\end{figure}

Figures \ref{fig:mainU} and \ref{fig:mainB} show exclusion plots 
constructed from total rates,
and from a spectral (binned) analysis of the simulated data, respectively.
For the binned case, 8 energy-bins and 8 R-bins of equal size are considered. Note that, for the
binned case, the corresponding
background in the detector is scaled accordingly (it grows radially as $R^2$).
Results for three ion sources are shown:
the $^{8}$Li-$16$ MeV, the $^{8}$Li-$13$ MeV and $^{6}$He cases.
The low $Q$-value of the helium ions clearly hinders the sensitivity of this setup, 
making it clearly inferior to the lithium ion source case.
Notice the slight difference between the two $^{8}$Li cases, which is only
marginally enhanced
for the binned case for large $\Delta m^2_{\textrm{new}}$($>$7~eV$^2$).
(Small) corrections from ions
decaying to the ground state of the daughter nucleus are thus expected to be
important only in the
large $\Delta m^2_{\textrm{new}}$ case.
The results of Figure \ref{fig:mainB} show the importance of an appropriate binning.

For comparison we have also included, in these figures, shaded regions corresponding to
the $99\%$ C.L. inclusion domains identified by the combination of data from the reactor neutrino
experiments and other sources as described in, and here adapted from, Fig.~8 of
Ref.~\cite{arXiv:1101.2755}.
One can see that the proposal investigated here would allow to cover most of the active-sterile oscillation
parameter region.
On the other hand we recall that the presented exclusion curves have the
following simple
physical meaning: an actual measurement lying inside the curve (to the upper-right of
the curve) represents
definite evidence in favour of the corresponding hypothesis, in our case, active
neutrinos oscillating into sterile ones. In this manner, from Figure
\ref{fig:mainB} one sees that the shaded region is out of reach if one uses 10$^{13}~^6$He/s; whereas
using 10$^{13}~^8$Li/s one can cover around 70$\%$-75$\%$ of the currently allowed region.

\begin{figure}[!]
\begin{center}
\epsfxsize=7.5cm\epsffile{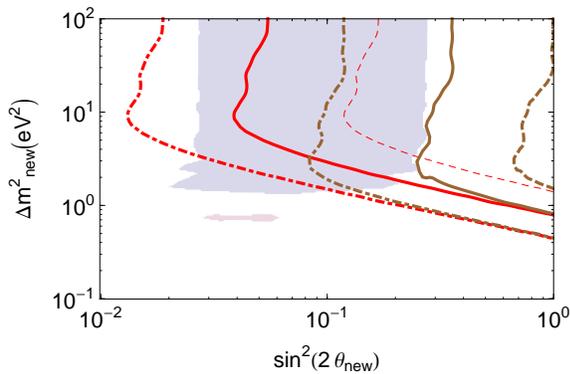}
\end{center}
\caption{Exclusion plots with binned analysis of the simulated data, obtained by varying
the ion intensity: $10^{14}$ ions/s (dashed-dotted), $10^{13}$ ions/s (solid) and
$10^{12}$ ions/s (dashed).
The red lines correspond to the source of $^{8}$Li ions
decaying mainly to the first excited state of the daughter nucleus (maximal
neutrino energy 13 MeV); while
the brown lines correspond to the source of $^{6}$He ions.
The contours shown are for a C.L. of 99\% (2 d.o.f.).
For comparison the shaded region represents the 99\% C.L.
inclusion domain, given by the combination of reactor neutrino
experiments and other sources (adapted from Fig. 8 of Ref.
\cite{arXiv:1101.2755}).}
\label{fig:Intensity}
\end{figure}

We would like to discuss now the impact of the chosen ion intensities
on the setup performance. Figure \ref{fig:Intensity} shows how the exclusion plots (and the coverage
of the allowed region) changes when varying the ion intensity. In particular,
the physics potential, relative to our main setup with
$10^{13}$ ions/s, is seen to diminish (increase) by changing the intensity in one order of
magnitude. This speaks
of the high level of influence, that achieving good ion production levels
nearby future radioactive ion beam facilities have, upon these
type of experimental searches.

\begin{figure}[!]
\begin{center}
\epsfxsize=7.5cm\epsffile{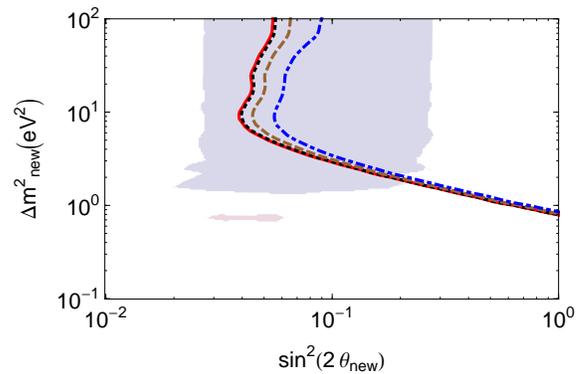}
\end{center}
\caption{Exclusion plots with binned analysis of the simulated data, obtained for the $10^{13}$ ions/s with $^{8}$Li ions
decaying mainly to a 13 MeV excited state of the daughter nucleus.
The figure shows the impact on the exclusion curves of different levels of systematic errors, namely 1\% (solid, red), 2\%  (dotted, black),
5\% (dashed, brown) and 10\% (dashed-dotted, blue).  
The contours shown are for a C.L. of 99\% (2 d.o.f.).
For comparison the shaded region represents the 99\% C.L.
inclusion domain, given by the combination of reactor neutrino
experiments and other sources (adapted from Fig. 8 of Ref.
\cite{arXiv:1101.2755}).}
\label{fig:sys1}
\end{figure}

\begin{figure}[!]
\begin{center}
\epsfxsize=7.5cm\epsffile{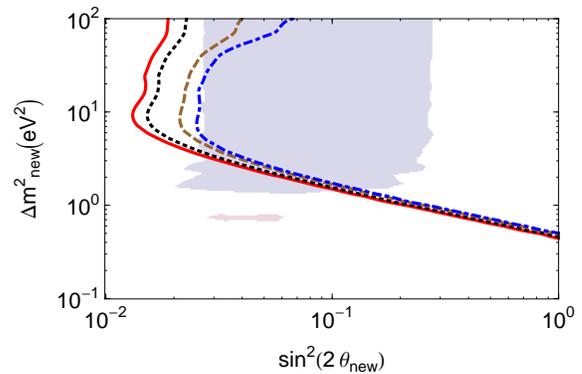}
\end{center}
\caption{Exclusion plots with binned analysis of the simulated data, obtained for the $10^{14}$ ions/s with $^{8}$Li ions
decaying mainly to a 13 MeV excited state of the daughter nucleus.
The figure shows the impact on the exclusion curves of different levels of systematic errors, namely 1\% (solid, red), 2\%  (dotted, black),
5\% (dashed, brown) and 10\% (dashed-dotted, blue).  
The contours shown are for a C.L. of 99\% (2 d.o.f.).
For comparison the shaded region represents the 99\% C.L.
inclusion domain, given by the combination of reactor neutrino
experiments and other sources (adapted from Fig. 8 of Ref.
\cite{arXiv:1101.2755}).}
\label{fig:sys2}
\end{figure}

Finally, the sensitivity of the proposed experiment might depend upon the
achieved systematic errors. To show their impact, we present exclusion curves based upon a binned analysis for sterile
neutrino mixing parameters, for different levels of systematic errors. Figures \ref{fig:sys1} and \ref{fig:sys2} 
show the impact of 1\%, 2\%, 5\% and 10\% systematic error on the exclusion curves, for $10^{13}$ and $10^{14}$ $^{8}$Li/s, respectively. 
One can see the important impact that reaching low systematic errors has, especially for large $\Delta m^2$ and small mixing angle.

\vspace{.15cm}


\section{Conclusions} \label{sec:conclusions}
\noindent
Future intense radioactive ion beam facilities can offer an unique opportunity to
perform searches for beyond the Standard Model physics, using low energy neutrino fluxes from
beta-decaying ions. Here we consider two configurations, where either the ions are stored in a storage ring, or they are sent into a target inside a spherical detector, filled in either with a noble liquid,
or with a scintillator (with the addition of Gadolinium).
Our results show that, 
as long as
the ions are not boosted, the spherical geometry scenario gives better results than the storage ring one.
We have presented predictions for the expected events associated to a coherent
neutrino-nucleus scattering measurement. The realization of such an experiment
heavily depends on the achievement of large ion intensities and reaching challenging low energy nuclear recoils. The second option
considered is a sterile neutrino search, that can be performed using electron
anti-neutrino detection through inverse beta-decay in a scintillator. 
We have presented exclusion plots obtained 
from total rates and from analysis including spectral information
(with binning in neutrino energy and
distance within the detector) of the simulated data.
In particular, the binned analysis gives interesting results for ion intensities
achievable at future radioactive ion beam facilities, like e.g. the EURISOL facility.  Clearly the ion intensities achievable at such facilities are
lower than the MCi radioactive source considered in proposals like the one in Ref.\cite{arXiv:1107.2335}.
However radioactive ion beam facilities offer the possibility to dispose of
radioactive ions with different Q-values, allowing to cover different regions 
of the oscillation parameters.
With our spherical setup, one can probe large squared-mass differences and rather small mixing
angle values, associated with one sterile neutrino, in the 3+1 oscillation
framework. In particular, with the kind of setup we consider here, one could confirm/rule out
the sterile neutrino hypothesis, as a possible explanation of the currently
debated reactor neutrino anomaly, and cover most of the corresponding parameter
space region.

\vspace{.3cm}


\begin{acknowledgments}

\noindent
We are grateful to Guido Drexlin for important discussions on background issues, for his encouragement in pursuing this project and for his careful reading of this manuscript.
We would like to thank  Klaus Blaum and Manfred Grieser for providing us with
the TSR characteristics, Kate Scholberg for information concerning backgrounds
and the CLEAR proposal, Thierry Stora for the ion intensities at HIE-ISOLDE,
Mauro Mezzetto for useful discussions on sensitivity issues.
C.E-H. acknowledges the support of IPN Orsay and in part by the Grant Generalitat Valenciana VALi+d,
PROMETEO 2008/004 and FPA 2008/02878 of Spanish Ministry MICINN.

\end{acknowledgments}

\vspace{.5cm}

\noindent
Once completed this manuscript, the authors have discovered
Ref.\cite{Agarwalla:2010gd} which presents an overlap with the present work.



\end{document}